\documentclass[%
 reprint,
 amsmath,amssymb,
 aps,showkeys
]{revtex4-2}

\usepackage{graphicx}
\usepackage{dcolumn}
\usepackage{bm}
\usepackage{multirow}
\usepackage[english]{babel}

\newcommand{\figref}[1]{Fig.~\ref{#1}}
\newcommand{\tabref}[1]{Table~\ref{#1}}

\renewcommand{\eqref}[1]{Eq.~(\ref{#1})}

\makeatletter
\def\bbl@set@language#1{%
  \edef\languagename{%
    \ifnum\escapechar=\expandafter`\string#1\@empty
    \else\string#1\@empty\fi}%
  \@ifundefined{babel@language@alias@\languagename}{}{%
    \edef\languagename{\@nameuse{babel@language@alias@\languagename}}%
  }%
  \select@language{\languagename}%
  \expandafter\ifx\csname date\languagename\endcsname\relax\else
    \if@filesw
      \protected@write\@auxout{}{\string\select@language{\languagename}}%
      \bbl@for\bbl@tempa\BabelContentsFiles{%
        \addtocontents{\bbl@tempa}{\xstring\select@language{\languagename}}}%
      \bbl@usehooks{write}{}%
    \fi
  \fi}
\newcommand{\DeclareLanguageAlias}[2]{%
  \global\@namedef{babel@language@alias@#1}{#2}%
}
\makeatother

\DeclareLanguageAlias{en}{english}
\DeclareLanguageAlias{eng}{english}
\DeclareLanguageAlias{en-US}{english}
\DeclareLanguageAlias{en-CA}{english}

\begin{document}

\preprint{APS/123-QED}

\title{High School Enrolment Choices --- Understanding the STEM Gender Gap}

\author{Eamonn Corrigan}
 \email{eamonn@uoguelph.ca}
\author{Martin Williams}
\affiliation{
 Department of Physics, University of Guelph, Guelph, Ontario, N1G 2W1, Canada
 }
 \author{Mary A. Wells}
\affiliation{
 Faculty of Engineering, University of Waterloo, Waterloo, Ontario, N2L 3G1, Canada
 }

\date{\today}

\begin{abstract}
\begin{description}
\begin{minipage}{5.55in}
    \item[Background]
Students’ high school decisions will always impact efforts to achieve gender parity in Science, Technology, Engineering, and Mathematics (STEM) at the university level and beyond. Without a comprehensive understanding of gendered disparities in high school course selection, it will be impossible to close completely the gender gap in many STEM disciplines. 

    \item[Results]
This study examines eleven years of detailed administrative data to determine gendered enrolment trends in university-stream secondary school STEM courses. Male and female enrolments for all publicly funded secondary schools across the province of Ontario ($N \approx 844$) were tracked from the 2007/08 academic year to 2017/18. The data reveal a clear trend of growing enrolment in STEM disciplines, with the increase in female students continuing their STEM education significantly outpacing males in almost all courses. However, these results also demonstrate the significant disparities that persist across STEM disciplines. The existing gender gap in physics remains large – in 2018, the median grade 12 physics class was only $36.5\pm0.05\%$ female – with virtually no progress having been made to close this gap. By tracking individual student cohorts, we also demonstrate a newly discovered result showing the continuation rate of male students in biology stream courses has experienced a precipitous drop-off. The proportion of male students continuing from grade 10 science to grade 12 biology two years later has seen an average yearly decline of $-0.44\pm0.08$ percentage points, potentially foreshadowing the emergence of another significant gender gap in STEM. 

    \item[Conclusions]
We suggest that researchers and educators cease treating STEM as a monolith when addressing gender disparities, as doing so obscures significant differences between disciplines. Future efforts, particularly those aimed to support women in STEM, must instead adopt a more targeted approach to ensure that they solve existing problems without creating new ones.
\end{minipage}
\end{description}

\end{abstract}

\maketitle

\section{Introduction}\label{introduction}

Excellence in Science, Technology, Engineering, and Mathematics (STEM) is built on a diversity of ideas and people. Women have historically been underrepresented in STEM fields and increasing gender diversity is more than a moral imperative, it is essential to maximize innovation, creativity, and competitiveness in Ontario and across Canada. The research is clear, heterogeneous groups are better at problem-solving than groups lacking diversity \cite{phillips_how_2014}, while organizations with diverse workforces are more economically productive than those without \cite{hunt_delivering_2018}. Despite decades of focused effort by multiple agencies to promote women in STEM (for a list of some initiatives in Canada see Canada STEM) yet continue to see a significant under-representation of women enrolled in many university STEM programs. Progress to improve outcomes at the undergraduate level is significantly constrained by enrolments in grade 12 high school STEM courses; to apply and gain entry to almost any undergraduate STEM program across Canada, students must complete several grade 12 STEM courses that satisfy mandatory admission prerequisites. For example, almost all engineering programs across Canada require having completed grade 12 Physics and Chemistry, as well as grade 12 Functions and Calculus. If we hope to disrupt our future talent pool and ensure an increased representation of women, we need to understand better the historical enrolment patterns in high school science courses across Ontario and how these differ between genders.

\subsection{Gender Disparities in
STEM}\label{gender-disparities-in-stem}

For decades, the gender disparity in STEM, which has historically led to
the underrepresentation of women, has been an area of extensive research
and intervention (see e.g., \cite{rossi_women_1965, brown_review_1966, eagly_gender_1984, keller_reflections_1985, stadler_boys_2000, porter_philosophers_2003, hazari_views_2005, han_having_2007, lock_physics_2013, bian_gender_2017, cheryan_why_2017, codiroli_mcmaster_who_2017, nserc_women_2017, breda_girls_2019}. Currently, we have a broader and somewhat more complete grasp of this issue and its root causes. As a result, tremendous progress has been made to achieving gender parity in STEM as a whole. In 2020/21, Ontario universities reported that over 48\% of STEM majors were female \cite{robertson_partnering_2022}, up from 43.6\% in 2010 \cite{wall_persistence_2019}, but growth has not been uniform across all STEM disciplines. Since the mid-1990s, female enrolment in STEM undergraduate programs like engineering and physics has plateaued at around 20\% \cite{wells_closing_2018}. This difference is even more pronounced in the workplace, where only 13\% of licensed Canadian engineers are female \cite{engineers_canada_engineers_2020}. In Ontario, the Canadian province with the largest number of licensed professional engineers, women make up 12\% of licensed professional engineers \cite{engineers_canada_engineering_2015}. Similarly, a volunteer-based demographic survey undertaken by the Canadian Association of Physicists found that only 35\% of physics faculty did not identify as men \cite{smolina_can_2021}. However, due to limitations of the survey's snowball distribution method, these results almost certainly over-represent these numbers. In contrast, the 2015 ratio of female to male students majoring in the physical/life sciences was 1.3:1 \cite{nserc_women_2017}, similar to the ratio found in Canadian medical schools where 63\% of students are women \cite{glauser_rise_2018}. In general, the proportion of Canadian undergraduate women who pursue a degree in STEM is now higher compared with the proportion of undergraduate men, but only for ``less math-intensive STEM fields'' or health sciences \cite{chan_gender_2021}. In mathematics and computer science, the percentage of female undergraduates is below 30\% \cite{government_of_canada_postsecondary_2021, nserc_women_2017}.

These trends are mirrored within STEM subdisciplines. In undergraduate
engineering programs such as biomedical or bioresources engineering and
environmental engineering, the participation of men and women is close
to parity \cite{gibbons_female_2019}. In contrast, fewer than 20\% of students in
mechanical, electrical, and computer engineering -- disciplines that are
viewed as being more closely related to physics -- are women \cite{gibbons_female_2019}. Overall, the increase in women's participation in STEM fields has not been uniform.

\subsection{Causes of Underrepresentation }

There are a multitude of factors that contribute to unequal
participation between genders. In a 2017 review, Cheryan et al. examined
the most common explanations presented in the literature to evaluate the
available evidence and understand the gender gaps between the STEM
disciplines \cite{cheryan_why_2017}. They identify three broad categories
supported by research: insufficient early experience; gender gaps in
self-efficacy; and the masculine culture of some STEM fields, i.e.,
stereotypes about who participates in a field, assumptions about a
woman's ability to succeed, as well as the lack of effective female role
models \cite{cheryan_why_2017}.

These stereotypes regarding women and STEM develop relatively early in
childhood. In one study, teachers were shown to view fictional
8-year-old students as less academically capable in physics and taught
them less scientific material when they thought the student was a girl
\cite{newall_science_2018}. Stereotypes about brilliance and intellectual
ability have also been shown to appear in children as young as six \cite{bian_gender_2017}. A perceived necessity for brilliance to succeed in a given academic field has subsequently been shown to promote masculine
cultures within those fields \cite{vial_emphasis_2022}, reinforcing these
gendered stereotypes about who belongs in STEM.

Gendered attitudes and preconceptions of STEM fields can also have a
significant effect. For instance, women who interact with a computer
science major who conforms to previously held stereotypes, e.g., nerdy,
socially awkward etc., are less likely to want to pursue a major in
computer science \cite{cheryan_enduring_2013}. Women compared with men have
also been shown to be more interested in people-oriented vs.
thing-oriented occupations \cite{lippa_gender_2010}, a disposition which helps
explain the participation gaps in the social or life sciences compared
with engineering and physics \cite{su_all_2015, yang_gender_2015}.
Similarly, a study of high school students in the United Kingdom found
female students, on average, lack interest in technical details while
having the desire to make a positive world impact with their work; both
these attributes were negatively correlated with the students'
intentions to continue pursuing physics \cite{mujtaba_what_2013} while
potentially explaining women's growing engagement with the biological or
life sciences.

\subsection{The Intent of this Work}\label{the-intent-of-this-work}
This paper aims to address two gaps in the existing literature. First,
most research to date treats STEM as one homogenous discipline or only
focuses on a single subject. Typically, data describing the state of
women in STEM across Canada combine multiple subjects into broad
categories such as ``Physical and Life Sciences'' or ``Science and
Technology'' \cite{hango_gender_2013, nserc_women_2017}. As outlined above, these
groupings conceal important differences between STEM disciplines.
Consequently, this research examines trends in the various high school
science disciplines and mathematics independently.

Most work to date has also focused on post-secondary education, e.g., \cite{grunspan_males_2016, lichtenberger_predicting_2012, turnbull_leaky_2017, yang_gender_2015} or the labour force, e.g., \cite{canada_advancing_2021, engineers_canada_engineering_2015, engineers_canada_engineers_2020}. A study of 932 undergraduate physics students found that high school was the most influential time for female students choosing to pursue a degree in physics, even among those who originally had no intention of studying STEM \cite{hazari_importance_2017}. Other research examining a cohort
of Ontario students transitioning from high school to university found
that students' STEM readiness, i.e., having taken the necessary STEM
courses in high school, accounted for 84\% of the gender gap seen in
undergraduate STEM programs \cite{card_high_2021}. To effectively
address this issue on a broader scale, it is paramount to develop a
better understanding of gendered enrolment patterns in high school.

For this work, we have conducted a case study on gendered STEM enrolment
in Ontario secondary schools. Ontario was chosen for its large
population size (15 million inhabitants; 38\(\%\) of Canada's
population) and its diverse population as measured by various
demographics such as racial or ethnic, socioeconomic, and a mix of urban
and rural geographies, among others \cite{government_of_ontario_census_2022, government_of_ontario_census_2022-1, government_of_ontario_census_2022-2, government_of_ontario_census_2022-3}.

There is a growing body of research that emphasizes the importance of
adopting an intersectional perspective when examining gender gaps in
STEM, including the interactions of race or socioeconomic status. For
example, research on secondary school students in the US found that
African-American women hold weaker gendered stereotypes about STEM
participation compared to their European counterparts, with a much
smaller disparity observed between men \cite{obrien_ethnic_2015}. Moreover,
a longitudinal study looking at the academic outcomes of eight grade
students found larger racial and gender disparities among students from
higher socioeconomic backgrounds compared to those within lower
socioeconomic status groups \cite{becares_understanding_2015}. This finding
underscores how economic factors can further exacerbate gendered or
race-based stereotypes. Finally, an intersectional examination of
participation in Advanced Placement Physics courses in high school
reported the largest gender gap in participation for women from
traditionally underrepresented ethnic minorities \cite{krakehl_intersectional_2021}.

While our study did not allow for a direct exploration of these nuanced
intersectional effects, our decision to focus on Ontario aims to
identify broader, general trends across a diverse range of school
populations. Through this work, we can begin to better understand
gendered differences in high school STEM across Canada, laying the
groundwork for more detailed intersectional analysis in the future.

\section{Methods}\label{methods}

\subsection{Ontario Ministry of Education
Data}\label{ontario-ministry-of-education-data}

\begin{figure*}
\includegraphics[width = 0.7\linewidth]{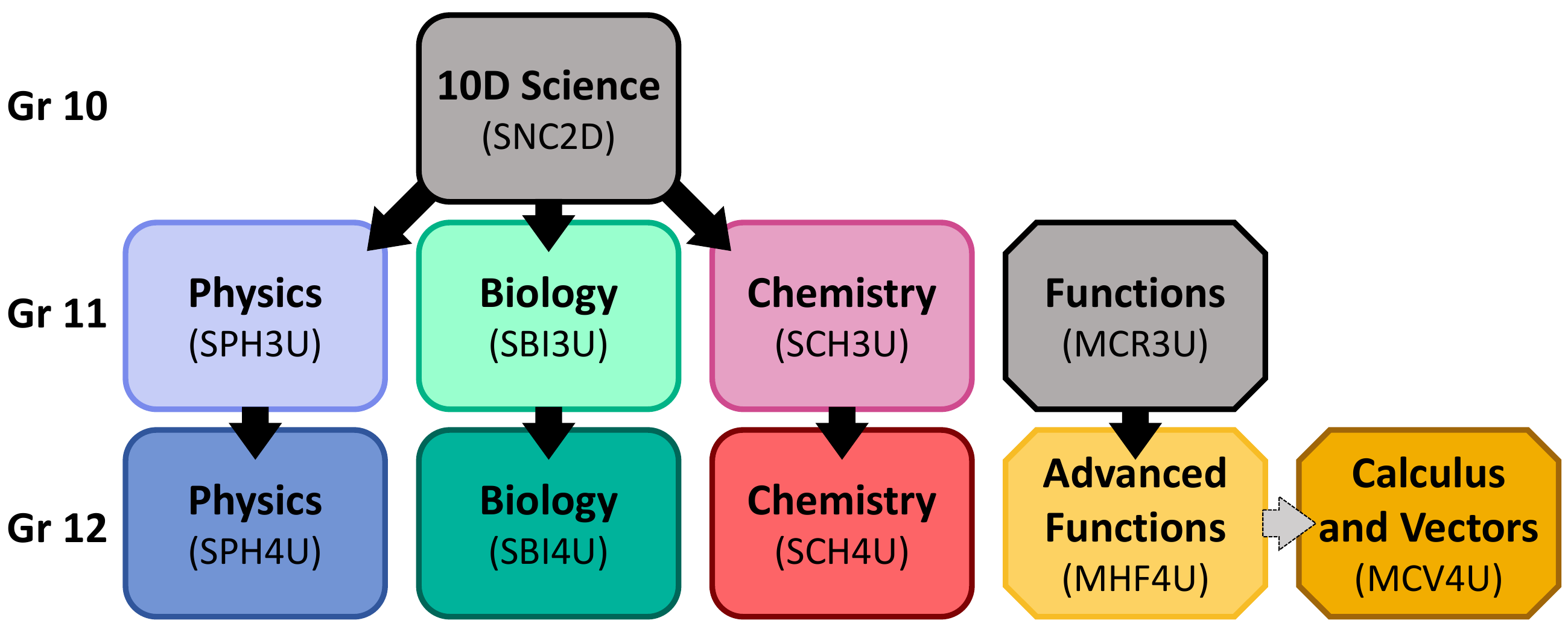}
\caption{Course pathway for Ontario schools showing all university
track, senior STEM courses included in this analysis. The top row is
grade 10, the middle grade 11, and the bottom grade 12. Courses in grey
represent the highest-level course required for graduation. All other
courses are colour coded by stream with blue for physics, green for
biology, red for chemistry, and yellow for mathematics. Black arrows
indicate a prerequisite transition and a greyed arrow with a dotted
outline indicates a course transition which can be taken as either a
pre-requisite or co-requisite.}
\label{fig:Courses}
\end{figure*}

We obtained the data used in this work from the Ontario Ministry of Education (OME). The dataset includes the total number of male and female \footnote{It is important to acknowledge this dataset only contains information about student sex, not their gender identity. This is the best available proxy, which is accurate for the vast majority of students, but not all.} students who enrolled in each university-track science and math course from grade 10 to grade 12, for every public secondary school in the province. The exact coursesbincluded are shown in \figref{fig:Courses}. The data span eleven academic years,
from 2007/08 to 2017/18.

For this date range, between 92.9-94.5\% of students in Ontario attended
a publicly funded school while almost all other students attended
private institutions \cite{government_of_canada_number_2021}. Analysis by
Statistics Canada found higher levels of academic achievement for
private school students including increased scores on standardized tests
and a higher likelihood to obtain a post-secondary degree. These
differences, however, were no longer significant once controlling for
socioeconomic characteristics and peer effects, finding no fundamental
difference in the students themselves \cite{frenette_postsecondary_2017}. Thus,
this sample of only public schools is likely representative of the
province as a whole.

To further protect student confidentiality, the OME chose to suppress all
low enrolment counts with fewer than 10 students to further protect
student privacy. Unless otherwise stated, schools with suppressed data
have been removed from all analyses in this paper.

This research project was reviewed by the University of Guelph Research
Ethics Board which deemed the project did not need full ethics approval
as we were only seeking to collect aggregate, not individual data for
analysis (REB \#19-06-015). The research department at the OME
subsequently reviewed the project, coming to the same conclusion.

\subsection{Statistical Analysis
Methods}\label{statistical-analysis-methods}

To characterize gendered enrolment trends in senior-level STEM courses,
four different analyses were conducted. This included 1) calculating
total enrolment in all senior STEM courses across the province and
measuring the change in enrolment over time, 2) calculating the median
female participation rate for each STEM course over time, 3) calculating
average Student Continuation Rate ($SCR$) for male and female students,
and 4) examining how $SCR$ for male and female students has been changing
over time. The details for each method are explained below.

\subsubsection{Total Enrolment in Science
Courses}\label{total-enrolment-in-science-courses}

The total number of male/female high school students enrolled in STEM
courses across Ontario was calculated for each of the eleven years of
available data, 2008-2018. To account for suppressed courses with
<10 students in a given year, several different imputation
methods were evaluated using a similar data set used in previous work
\cite{wells_closing_2018}. This second data set, which contained overlapping
years from 2007/08-2014/15, reported total enrolment rates for entire
school boards instead of individual schools. As each board is much
larger and was unlikely to have a total enrolment below ten, almost no
data were suppressed in this second data set, allowing us to evaluate
which imputation technique provided the most accurate estimates. It was
determined that imputing a value of 9 for each suppressed school most
closely matched the enrolment counts obtained from board-level data.
Since courses with extremely low enrolment rates are unlikely to be
offered due to limited resources, it is reasonable to assume that the
true value of the majority of suppressed cells is significantly closer
to ten than zero. Figure \ref{fig:Tot_Enrol} plots the total enrolments of male and female
students in all STEM courses between grades 10 and 12 using this
imputation method to fill in the missing data from suppressed schools.

To quantify the average change in total enrolment over time, a simple
linear regression to model total enrolment (male and female combined)
was calculated using time as a predictor variable. Figure \ref{fig:Tot_Enrol}  includes the
resulting regression lines and estimates for the average annual change
in enrolment.

\subsubsection{Median Female Enrolment Over
Time}\label{median-female-enrolment-over-time}

The percentage of students who were female and enrolled in each STEM
course at each school from grade 10 to grade 12 was calculated over the
eleven years for which data was available. The results were then
averaged over all schools for each year to plot the median female
proportion in all senior STEM courses vs. time (\figref{fig:F_Prop}). In addition
to the visual representation, we sought to quantify the change in median
female participation over time. A weighted least squares linear
regression was performed comparing the median female proportion versus
time for each course. The weights used were \(1/SD^{2}\) where \(SD\) is
the sample standard deviation of the median female proportion.

\subsubsection{Average Student Continuation
Rates}\label{average-student-continuation-rates}

To track the progression of students through the high school STEM
courses, each school's enrolment data was grouped into
three-year cohorts. Then, these linked cohorts were used to calculate
\emph{Student Continuation Rates} ($SCR$). For science courses, $SCR$ is
defined as the number of male and female students enrolled in each of
the grade 11 courses divided by the number of male and female students
who were enrolled in grade 10 science the previous year. Grade 12
courses were calculated the same way but using grade 10 enrolment from
two years prior. In both cases, grade 10 science was used as the
reference category as it is the highest-level science course required
for graduation. For mathematics courses, $SCR$ for grade 12 Functions and
grade 12 Calculus were defined as the ratio of male and female students
enrolled in each course divided by the enrolment in grade 11 Functions
from one year prior. This was selected as reference since grade 11 is
the highest-level math course required for graduation.

It should be noted that this method does not perfectly describe student
cohorts -- some students will move between schools and others may not
take their courses in a linear progression -- but this should still
provide a good approximation for the vast majority of students.

To understand the continuation or attrition of students between the last
mandatory STEM course and grade 12, $SCR$ was calculated for each of the
four STEM streams for both male and female students. This was then
averaged across all years of data to quantify the average loss of
potential students during this pivotal period.

\subsubsection{Changes in Male and Female Student Continuation
Rates}\label{changes-in-male-and-female-student-continuation-rates}

The previous 3 analyses helped to characterize how male and female
enrolments have or have not been changing over time, but do not explain
the underlying mechanisms driving these trends. For example, the median
proportion of female students could be increasing because female
participation is rising, male participation is falling, etc. To
determine the mechanism(s) underlying these trends the average yearly
change in $SCR$ \((\Delta SCR)\) for different courses was measured using
a mixed effects linear regression model. The model includes fixed
effects variables for year \(\left( X_{1} \right)\), and student sex
\(\left( X_{2} \right)\), as well as an interaction term between student
sex and year \((X_{1}X_{2})\). A random effect variable which accounts
for fluctuations in \(\Delta SCR\) between schools was also included.
The resulting model has the following form:

\begin{align}
\label{eq:reg}
    Y_{i} = \beta_{0} + \beta_{1}X_{1} + \beta_{2}X_{2} + \beta_{3}X_{1}X_{2} + S_{i}X_{1} + \epsilon.
\end{align}

Here, \(Y_{i}\) is the mean Student Continuation Rate in one of the
senior STEM courses at school \(i\). The fixed effect regression
estimates for year and student sex are represented by \(\beta_{1}\) and
\(\beta_{2}\) respectively, while the interaction term regression
estimate is \(\beta_{3}\). The random effect for school ID is
represented by \(S_{i}\), i.e., the average variation in \(\Delta SCR\)
from the provincial average for school \(i\). Finally, \(\beta_{0}\) is
the estimated intercept.

For this analysis, we are only interested in measuring \(\Delta SCR\)
for both male and female students and if there is a statistically
significant difference in \(\Delta SCR\) between the sexes, i.e., the
slope estimates \(\beta_{1}\) and \(\beta_{3}\) and their \(p\)-values.
The average difference in \(SCR\) between sexes, while included in the
model, was not the main focus as this was most previously examined by
calculating average $SCR$.

A \(t\)-test was performed to determine an associated \(p\)-value for
\(\beta_{1}\) and \(\beta_{3}\). The model was also calculated twice for
this analysis, once with the coding
$(X_{2} := \{male = 1, female = 0\})$ and once with the opposite.
Mathematically, these are equivalent -- the regression estimates for
\(\beta_{2}\) do not change except for a switch in sign, but this
enabled us to run a \(t\)-test on the regression estimate \(\beta_{1}\)
when it represented \(\Delta SCR\) for both male and female students.
This regression model was run for all senior STEM courses in our
dataset.

\section{Results}\label{results}

\subsection{Total Enrolment in Science Courses }\label{total-enrolment-in-science-courses-1}

\begin{figure*}
\includegraphics[width = \linewidth]{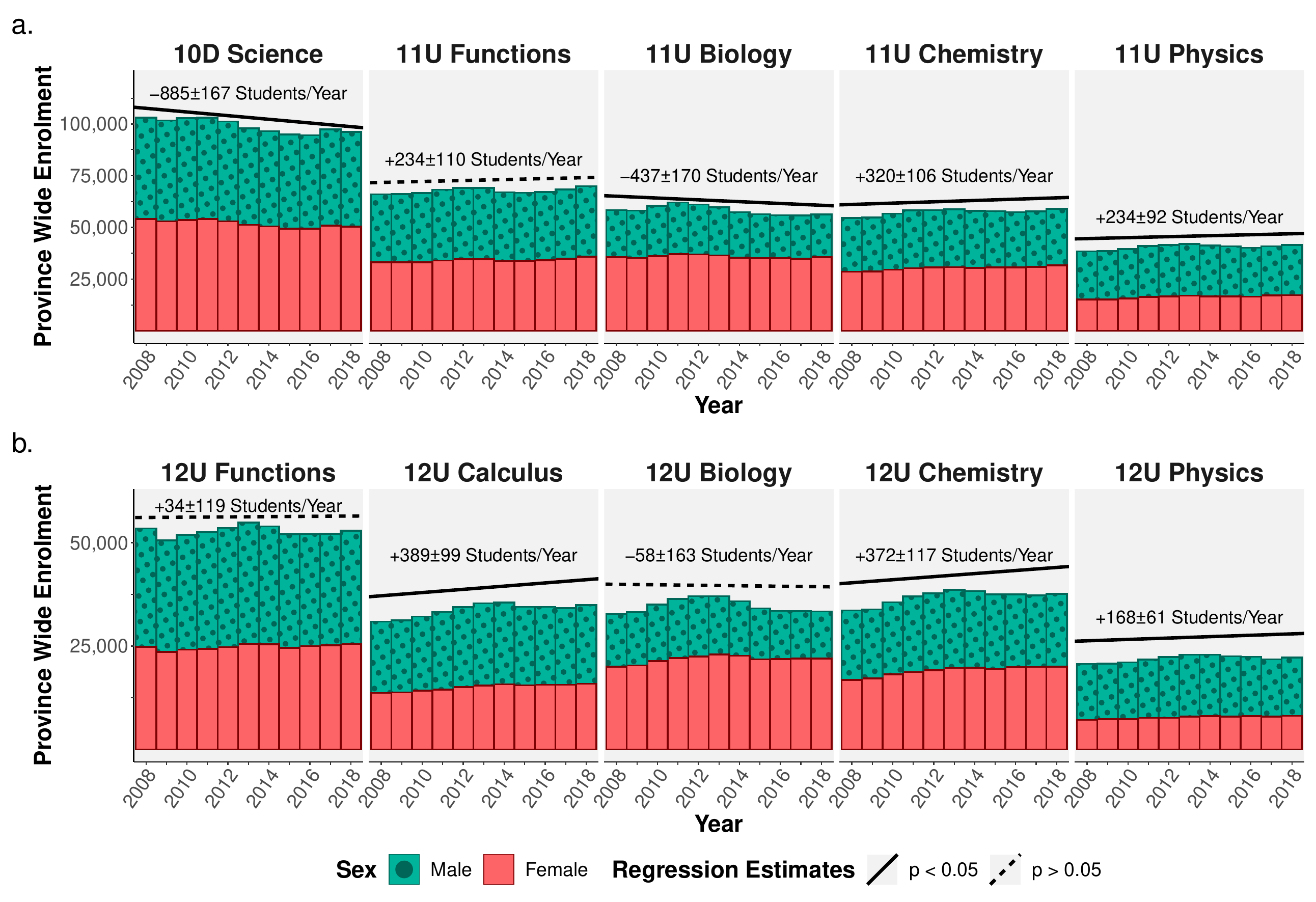}
\caption{Total enrolment in university track STEM courses from grade 10
through grade 12 for all secondary schools across Ontario. Sub-figure a.
shows grade 10 and grade 11 courses while sub-figure b. shows grade 12
courses. The lower, red, solid-coloured bars represent the total
enrolment of female students while the upper, green, patterned bars
represent the total enrolment of male students. The black lines depict
the slope estimate from a simple linear regression of total enrolment
vs. time. Solid lines indicate the regression estimates were
statistically significant \(p < 0.05)\), while solid lines were not. The
text above each line shows the regression estimate for the average
yearly change in total enrolment \(\pm\) the standard error. Sub-figures
a and b were plotted so the scale of a. is exactly twice the scale of b.
to ease comparison while still selecting scales which are most
appropriate for each.}
\label{fig:Tot_Enrol}
\end{figure*}

Total enrolment for all senior STEM courses in our dataset along with
the regression estimates and associated standard errors quantifying
average yearly change in enrolment is shown in \figref{fig:Tot_Enrol}. The results of
our model show a decreasing trend in the total number of students
enrolled in grade 10 Science $(\beta = (- 885 \pm 167, p = 0.00049))$,
which is likely linked to the known decline in total enrolment across
Ontario secondary schools \cite{pollock_principals_2016, robertson_declining_2014}. Despite
this overall drop in enrolment, most courses in Ontario were found to
have either stable or increasing total enrolment, implying a growing
interest in STEM. The sole exception is 11U Biology, the only course
other than grade 10 science which has seen a large, yearly decrease in
total student enrolment.

\figref{fig:Tot_Enrol} also demonstrates a stark contrast between the total enrolment
rates in physics stream courses compared with all other STEM
disciplines. Both grade 11 and grade 12 physics have the lowest total
enrolment in their respective grades, severely limiting the
undergraduate talent pool for physics and engineering. Total enrolment
in 12U Calculus is comparable to enrolment in 12U Biology and Chemistry,
challenging the notion that students\textquotesingle{} inability to
perform advanced mathematics is a substantial deterrent to enrolling in
physics. In 12U Calculus, the ratio of male to female students also
appears significantly more balanced than in 12U Physics which is largely
male-dominated. In contrast, there appears to be a greater proportion of
female students in 12U Biology than male students.

\subsection{Median Female Enrolment Over Time}\label{median-female-enrolment-over-time-1}

\begin{figure*}
\includegraphics[width = 0.7\linewidth]{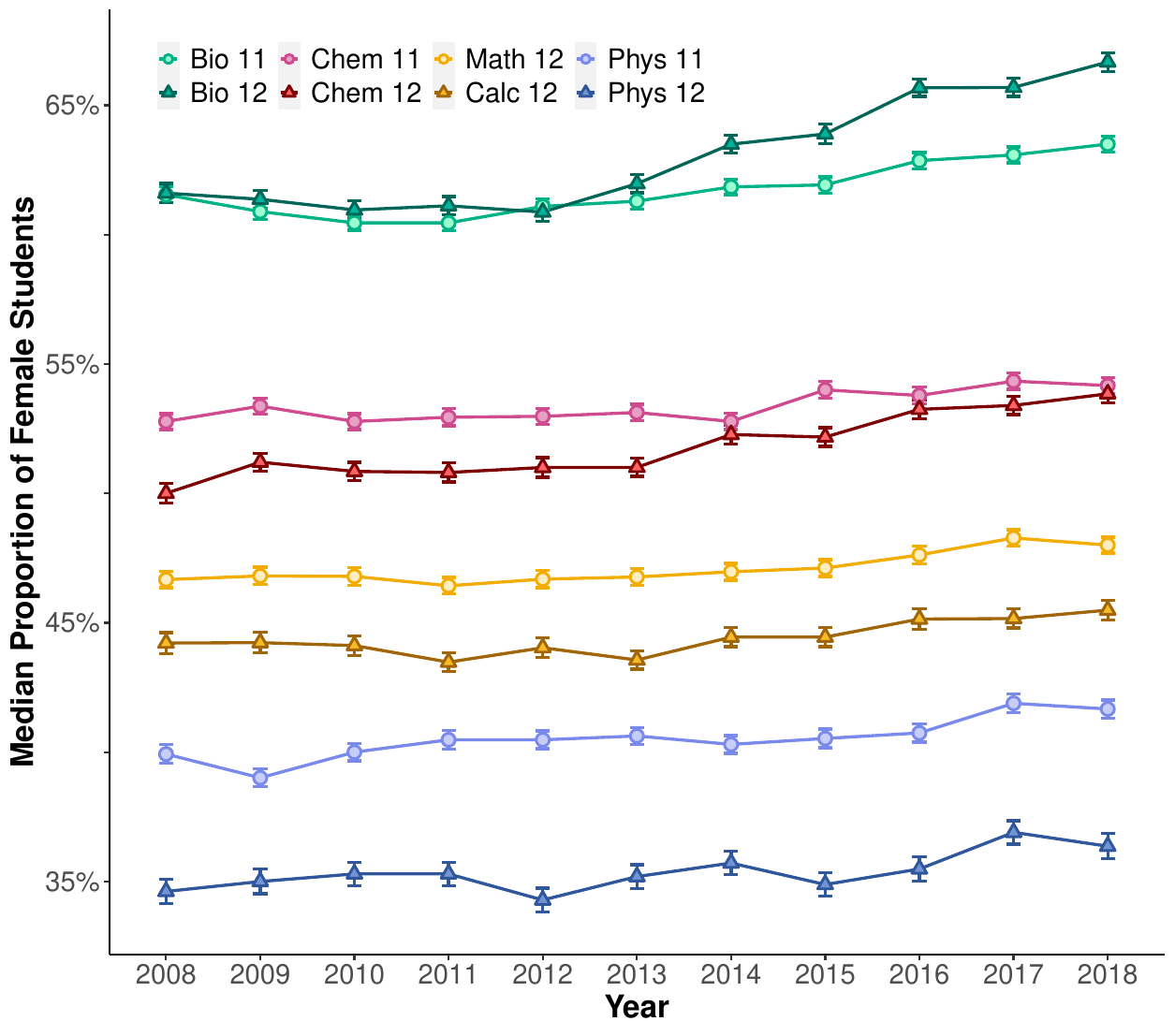}
\caption{The median proportion of female students enrolled in science
courses averaged across all Ontario secondary schools. The date on the
horizontal axis represents the second half of each academic year, e.g.,
2008 is the academic year spanning from September 2007 until June 2008.
Colour indicates the STEM stream: green for Biology, red for Chemistry,
blue for Physics, and yellow for Mathematics. Shape and shade indicate
the level of each course relative to the last mandatory STEM course.
Light-coloured circles represent the first post-mandatory course (11U
for the science courses and 12U Functions for Mathematics);
dark-coloured triangles represent the second post-mandatory course
(grade 12U for the sciences courses and 12U Calculus for Mathematics).
The error bars shown depict standard error.}
\label{fig:F_Prop}
\end{figure*}

The calculated values of median female proportion over time a plotted in
\figref{fig:F_Prop}. Note, grade 10 Science and Grade 11 Functions are left off
\figref{fig:F_Prop} for readability. For all eleven years, grade 10 Science had a
median proportion of female students around 52.5\% (min:
\(52.3 \pm 0.2\%\); max: \(52.7 \pm 0.3\%\)) while Grade 11 Functions
was just over 50\% (min: \(50.0 \pm 0.3\%\); max: \(51.2 \pm 0.3\%\)).

The results presented in \figref{fig:F_Prop} show that Grade 11 and 12 Chemistry
both have median female proportions just above 50\%, consistent with the
female proportion seen in Grade 10 Science. The two mathematics courses
are the next closest to parity. Grade 12U Advanced Functions is
approximately 47\% female (min: \(46.4 \pm 0.3\%\); max:
\(48.3 \pm 0.3\%\)) while 12U Calculus and Vectors sit around 44\% (min:
\(43.5 \pm 0.4\%\); max: \(45.5 \pm 0.4\%\)). The participation rate of
female students in these courses is lower than that of male students,
but the gap is not enormous. This once again contradicts the argument
that the gender gap in physics is primarily linked to ability in
advanced mathematics.

In contrast to chemistry and mathematics, enrolment disparities between
male and female students are notably pronounced in biology and physics.
The former is predominantly female while the latter is predominantly
male, while the size of the disparity is also of roughly equal magnitude
in both courses. For example, in 2017/18, grade 11 Physics had a median
male proportion of \(58.3 \pm 0.4\%\) while grade 11 Biology was
\(63.5 \pm 0.3\%\) female. In the same year grade 12U Physics was
\(63.6\  \pm 0.5\%\) male and grade 12U Biology had a median female
proportion of \(66.7 \pm 0.4\%\) female.

Visually, median female proportions seen also appear to be fairly stable
over the eleven years of available data. Except for 12U Biology and 12U
Chemistry, most courses appear to show only small changes in the
proportion of female students relative to the size of the standard
errors. The results of our weighted least squares regressions, which
quantify the change in median female proportion over time, are presented
in \tabref{tab:prop_vs_time_reg}.

\begin{table*}
\caption{\label{tab:prop_vs_time_reg}
Linear Regression Estimates for change in median female proportion over time.}
\begin{ruledtabular}
\begin{tabular}{lddd}
Course & \multicolumn{1}{l}{Estimate} & \multicolumn{1}{l}{Standard Error} & \multicolumn{1}{l}{$p$-value} \\
\hline
10D Science & -0.013 & 0.013 & .339 \\
11U Biology \footnote[1]{Indicates the regression estimate was found to be statistically
significant \((p < .05)\)} & +0.272 & 0.054 & <.001 \\
12U Biology\footnotemark[1] & +0.594 & 0.0901 & <.001 \\
11U Chemistry\footnotemark[1] & +0.141 & 0.036 & .004 \\
12U Chemistry\footnotemark[1] & +0.352 & 0.045 & .001 \\
11U Physics\footnotemark[1] & +0.203 & 0.041 & <.001 \\
12U Physics\footnotemark[1] & +0.157 & 0.054 & .017 \\
11U Functions\footnotemark[1] & +0.086 & 0.046 & .093 \\
12U Functions\footnotemark[1] & +0.153 & 0.032 & .001 \\
12U Calculus\footnotemark[1] & +0.140 & 0.046 & .014 \
\end{tabular}
\end{ruledtabular}
\end{table*}

Grade 10 Science and Grade 11 Functions are the only courses found to
have no statistically significant change in median female proportion.
Since both courses fulfill graduation requirements, it is not surprising
that these courses appear more resistant to change. For grade 11
Functions, it should also be noted that the p-value is equal to only
\(0.09.\) While the evidence for a non-zero change in the median female
proportion is not as strong as in the other STEM courses, it would be a
mistake to claim with certainty there has been no change over time.

Of the remaining STEM courses, all show an \emph{increase} in the median
proportion of female students. The effect size, however, varies widely
between STEM courses. Both grade 12 mathematics courses show evidence of
a slow increase in the median female participation rate with regression
estimates around \(0.15\). This corresponds to a one percentage point
increase in the median female proportion every 6-7 years. Grade 11
Chemistry, Grade 11 Physics, and Grade 12 Physics also show comparably
small effect sizes (all regression estimates of \(\leq 0.2\)).

The slow rate of growth for 12U Physics stands in stark contrast to the
other 12U science courses. The average yearly change in median female
participation for 12U Biology is 3.8x greater than 12U Physics; for 12U
Chemistry the average yearly change is 2.2x greater. This means both
chemistry and biology, which already have median female proportions
above 50\%, are continuing to widen this gap -- particularly biology. In
addition, while there is evidence for a statistically significant
increase in the median female proportion of 12U Physics, the overall
effect has been minimal. Extrapolating from the 2018 level of
\(36.4 \pm 0.5\%\) and assuming a constant annual change of 0.16
percentage points, it would take 85 years -- until 2103 -- for grade 12U
Physics to reach a median female proportion of 50\%.

\subsection{Average Student Continuation
Rates}\label{average-student-continuation-rates-1}

\begin{figure*}
\includegraphics[width = 0.8\linewidth]{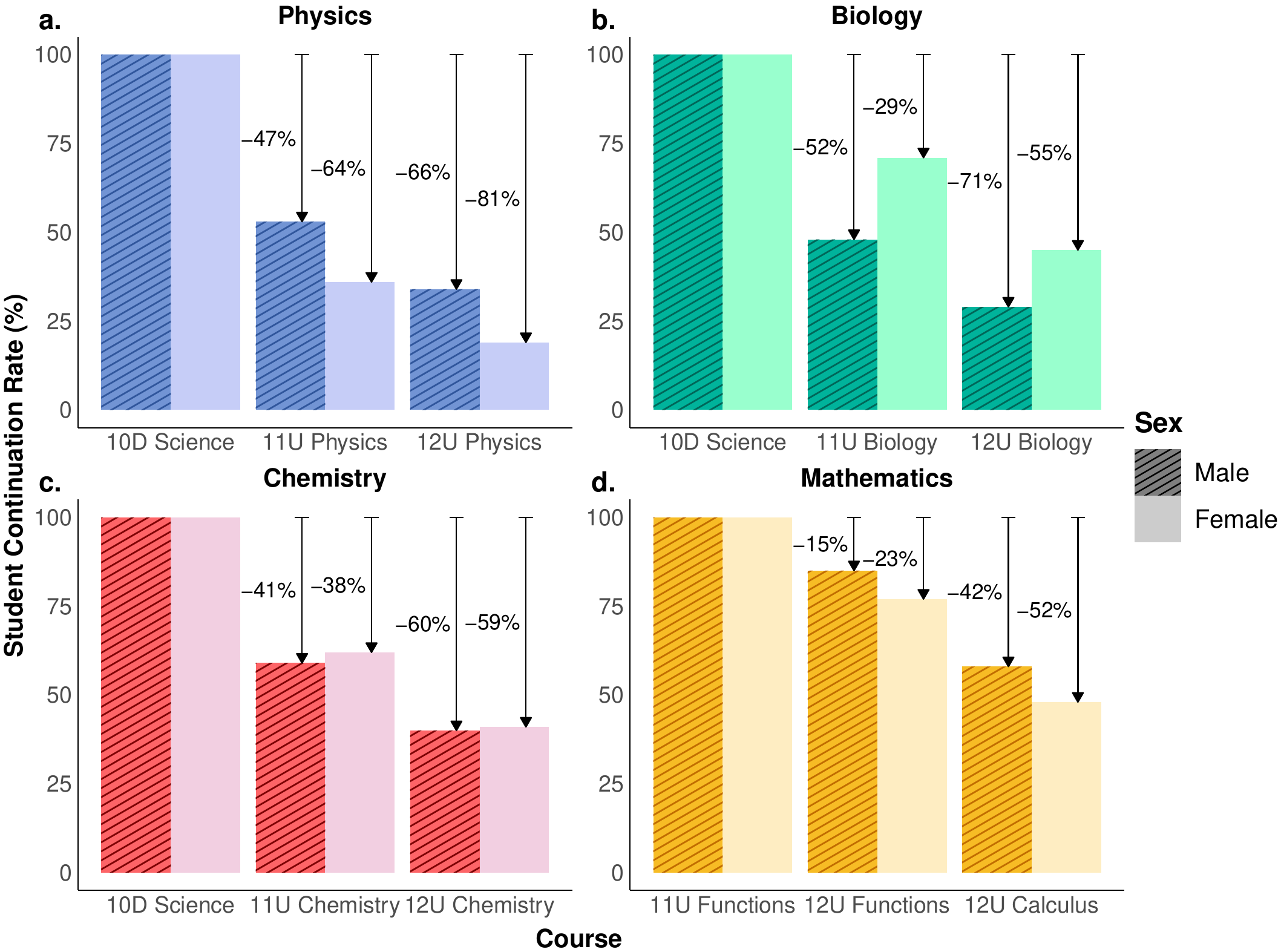}
\caption{Student Continuation Rate, averaged over all schools in
Ontario across all eleven years of available data. Student Continuation
Rate tracks student cohorts year to year in each school. Course colour
coding is consistent with \figref{fig:F_Prop}, where Sub-figure a. (Blue) shows
Physics stream courses, b. (Green) shows Biology stream courses, c.
(Red) represents Chemistry stream courses, and d. (Yellow) depicts
Mathematics courses. For all figures, the dark-coloured, patterned
columns represent male students while light-coloured, solid columns
represent female students. Arrows indicate student attrition rate, the
converse of student continuation rate, i.e., how many students are lost,
on average, since the last mandatory STEM course.}
\label{fig:Att_Plot}
\end{figure*}

Total enrolments and median female participation rate can only describe
the students who \emph{do} enrol in senior STEM courses -- a minority
of the student population. \figref{fig:Att_Plot} plots male and female students'
average $SCR$ for all four STEM streams, quantifying the levels of
attrition from mandatory STEM courses through to the end of high school.

The results clearly show that mathematics courses have higher
continuation rates than any of the sciences, especially Grade 12U
Advanced Functions. More than 77\% of female students and 85\% of male
students who take functions in grade 11 appear to continue to 12U
Functions the following year. This is likely due to the ubiquity of 12U
Functions as an admission requirement to numerous undergraduate
programs. Mathematics courses, as well as chemistry, also have the
smallest male/female gaps in $SCR$, matching the median female proportions
(\figref{fig:F_Prop}d).

Biology and physics possess the largest gendered gaps in $SCR$. Almost
twice as many male students do not continue from grade 10 Science to
grade 11 Biology compared with female students -- a 23 percentage point
gap in $SCR$. A larger proportion of these male students who take grade 11
do appear to remain until grade 12 as the gap in $SCR$ between male and
female students to sixteen percentage points for grade 12. Male and
female students in physics show comparable differences in $SCR$; the gap
in $SCR$ is seventeen percentage points in grade 11 and 15 percentage
points in grade 12. In addition, physics has the greatest overall loss
of potential students compared with any of the other three STEM
disciplines. The data shows that 66\% of male students and 81\% of
female students in grade 10 science will, on average, not continue to
grade 12 physics. The closest comparable loss is the 71\% student
attrition rate for male students continuing to grade 12 Biology.

\subsection{Changes in Male and Female Student Continuation
Rate}\label{changes-in-male-and-female-student-continuation-rate}

\begin{figure*}
\includegraphics[width = 0.7 \linewidth]{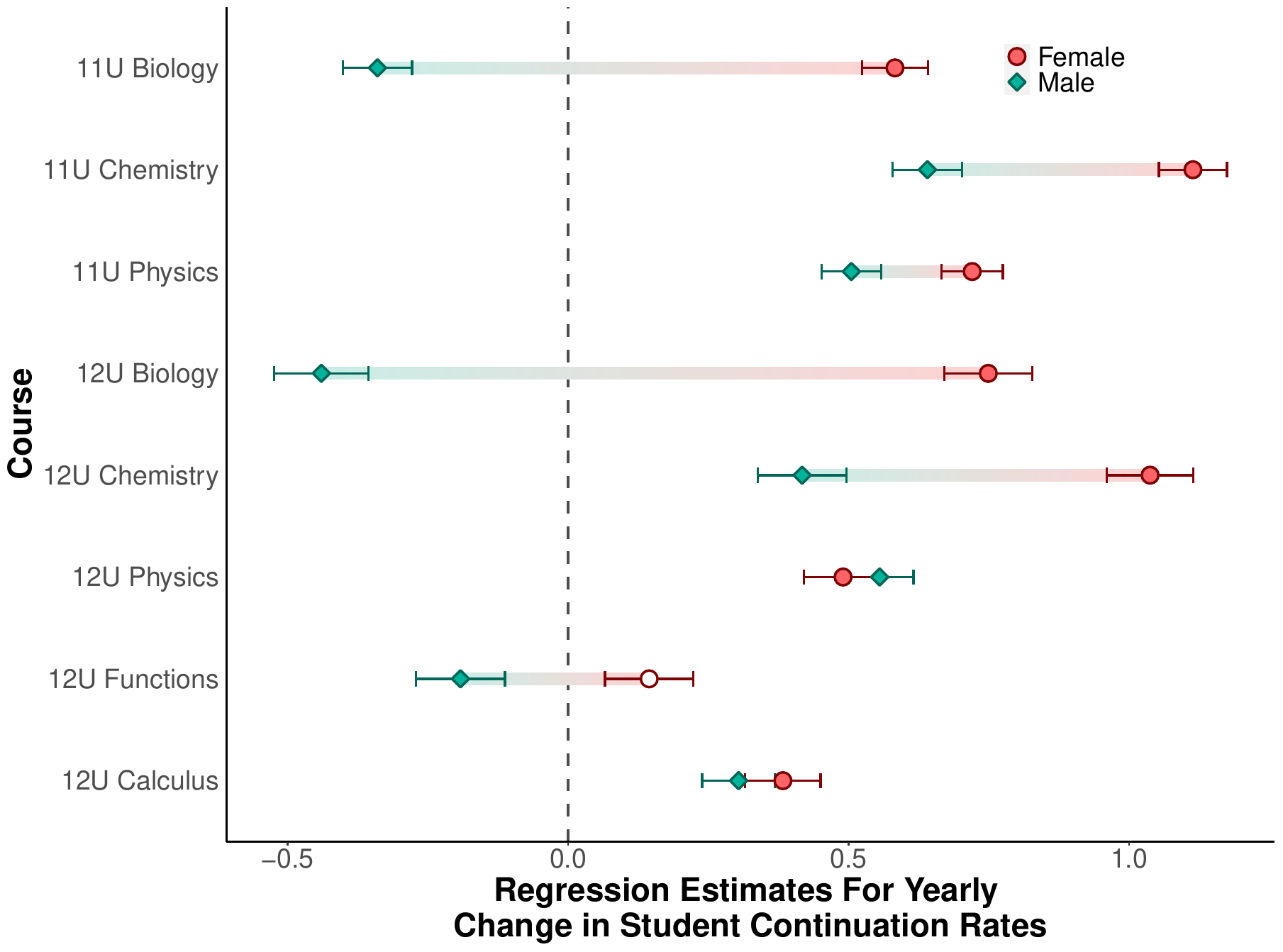}
\caption{Barbell plot of regression estimates modelling average change in
male/female $SCR$, per year, controlling for the random effect of
individual schools, including sex as an interaction term. Red circles
indicate female student enrolment and green diamonds male. Filled shapes
indicate the regression estimate was found to have a statistically
significant difference from zero ($p <0.5$); unfilled shapes
indicate the estimate was not found to be statistically different from
zero. The connecting lines between male/female students are coloured and
solid when the interaction term showed a statistically significant
\(\Delta SCR\) between male and female students ($p<0.5$);
points not connected with lines were found to have a \(\Delta SCR\)
which was not statistically different from zero.}
\label{fig:Delta_SCR}
\end{figure*}

The most important results of the mixed effects linear regression model
defined in \eqref{eq:reg} are plotted in \figref{fig:Delta_SCR}, i.e. \(\Delta SCR\) for
male and female students and the gap in \(\Delta SCR\) between the
sexes. A table with the full regression results from this analysis is
available in the appendix (\tabref{tab:delta_SCR_reg}).

These data show significant growth in $SCR$ for female students for almost
all STEM courses (p \textgreater{} 0.05; \tabref{tab:prop_vs_time_reg}). The sole exception
was grade 12 functions
\(\left( \beta_{1} = 0.14 \pm 0.8;p = 0.066 \right)\). While the t-test
conducted comparing the slope associated with \(\Delta SCR\) for grade
12 functions to zero did not have a p-value smaller than the widely used
cut-off value of 0.05, there is still some evidence to suggest $SCR$ for
female students increased in this course as well (\(p\  = \ 0.066)\).
These data show that the growth in the median proportion of female
students taking STEM courses (\tabref{tab:prop_vs_time_reg}) is primarily attributable to an
increase in the $SCR$ for female students. The rate of growth in the $SCR$
of female students has also been quite substantial. For example, the
regression estimates of \(\Delta SCR\) for female students in grades 11
and 12 Chemistry were found to be \(\Delta SCR_{11} = 1.11 \pm 0.06\)
percentage points per year and \(\Delta SCR_{12} = 1.04 \pm 0.08\)
percentage points per year respectively. For context, since
\(\sim 50,000\) female students take grade 10 Science every year, a
\(\Delta SCR \cong 1\) amounts to an extra five hundred female students
taking chemistry, compounding annually.

For male students, there has been similar, though smaller yearly growth
in $SCR$ for all STEM courses but three. 12U Functions
\((\beta_{1} = - 0.19 \pm 0.08;p = 0.015)\) as well as 11U
\(\left( \beta_{1} = - 0.34 \pm 0.06;p = 3.36 \times 10^{- 8} \right)\)
and 12U Biology
\(\left( \beta_{1} = - 0.44 \pm 0.08;p = 1.79 \times 10^{- 7} \right)\)
have all seen an average decline in male continuation over the eleven
years of available data.

When comparing between sexes, female students were found to have greater
estimates for \(\Delta SCR\) than male students in almost all STEM
courses. In only one subject, 12U Physics, male students had a greater
regression estimate for \(\Delta SCR\), but the interaction term was not
statistically significant \((p\  = \ 0.39),\) thus, there is
insufficient evidence to conclude that there is a genuine difference
between the sexes. Similarly, the interaction term for Calculus 12 was
not statistically significant \((p = 0.33)\), while all other courses
had non-zero differences in \(\Delta SCR\) between the sexes. The
magnitude of the gap in \(\Delta SCR\) between the sexes also varies
widely between the different STEM courses with the largest gap appearing
in biology stream courses. For example, the interaction term which
calculated the average difference in \(\Delta SCR\) between male and
female students was \(- 1.19 \pm 0.09\) percentage points per year. To
get a sense of scale for these estimates, we can interpolate by assuming
50,000 male and female students took grade 10 science every year. Over
the eleven years of data, assuming constant \(\Delta SCR\) as estimated
from the model, 29,040 fewer male students and 49,500 more female
students took 12U Biology between 2007 and 2018 compared with a
hypothetical scenario where \(\Delta SCR\) was equal to zero for both.

This explains the rapidly growing median female proportion seen in
\figref{fig:F_Prop}. The growth in the median percentage of female students is
driven from both directions as more female students are continuing their
biology education while more male students are choosing not to continue.
In contrast, the largest, positive estimates for \(\Delta SCR\) found
for male students appear in the chemistry and physics stream courses,
STEM subjects known to already have increased participation of males in
post-secondary and beyond \cite{cheryan_why_2017}.

\section{Discussion}\label{discussion}

\subsection{Growing Enrolment in STEM}\label{growing-enrolment-in-stem}

In this study we used enrolment data from the Ontario Ministry of
Education, separated by student sex, to quantify differences between
male and female students in university-track, high school STEM courses
from 2007-2018. We found clear evidence for a general trend of growing
engagement in STEM for both male and female students. Despite a total
decline in the province's study body, total enrolment in
nearly all senior stem courses remained flat or increased. The
continuation rate of students from STEM courses which are mandatory for
graduation to optional, senior-level courses has also increased in
almost all cases. For the eleven years of available data, the results
are clear: a larger share of high school students are choosing to study
STEM subjects at a senior secondary school level now compared with a
decade ago. This mirrors the trend seen in Canada's post-secondary
institutions. From 2010 to 2019, the total number of students studying
STEM increased by over 40\%, outpacing the increase in total
post-secondary enrolment \cite{mahboubi_knowledge_2022}.

There are many possible explanations for this growth in high school STEM
enrolment. As the labour force continues to specialize and advance, STEM
training is increasingly required to remain competitive in the job
market \cite{mahboubi_knowledge_2022}. As students, teachers, and families become more
aware of this growing demand, it is logical this would influence
students' course decisions. Students represented in this dataset have
also come of age during or after the 2008 recession. Previous work has
shown a strong positive correlation between students growing up during
periods of economic uncertainty and selecting majors with high potential
earnings like STEM \cite{blom_investment_2021}. Both the provincial and federal
governments have also implemented policies which both implicitly and
explicitly further this push for STEM education \cite{decoito_stem_2016, government_of_canada_government_2021}.

\subsection{Uneven Growth in Male and Female
Enrolment}\label{uneven-growth-in-male-and-female-enrolment}

While high school student enrolment in STEM has increased overall, a
closer look at the data reveals considerable variations between fields
and between the sexes. At the same grade level, overall enrolment in
physics courses is at least 30 percent lower than in biology, chemistry,
and mathematics courses. This arises from the extremely low continuation
rates of high school students from grade 10 Science: 66\% of male
students and 81\% of female students are lost by grade 12 physics. As
grade 12 physics is a prerequisite for nearly all university physics and
engineering programs in Canada, this represents the greatest loss of
potential talent throughout the education pipeline.

Physics also has one of the greatest gender gaps among high school STEM
courses. In 2018, the median percentage of female students enrolled in
grade 12 Physics courses was only \(36.3 \pm 0.05\%\). While there is
statistically significant evidence for a modest increase in the
proportion of female students enrolled in grade 12 physics as well as an
increase in the percentage of female students continuing from grade 10
science to grade 12 physics, both of these effects are minimal. As
discussed in the results section, it would take 85 years to close the
gender gap in 12U Physics at the current rate of change.

These statistics mirror those found in other settings. The American
Physical Society reported that from 2016-2020, only 22\% of
undergraduate degrees in physics or engineering went to women
\cite{aps_bachelors_2022}. In Ireland, an
analysis of the secondary to post-secondary transition found 24\% of
male students but only 7\% of female students studied physics as part of
their final Leaving Certificates \cite{delaney_understanding_2019}. The gap in
physics participation between male and female students is a well-known
and regularly discussed problem within the existing literature. However,
we have been unable to find any direct mention of the opposite problem
which we discovered for biology stream courses, though data do indicate
similar imbalances elsewhere. In the US, 63\% of
bachelor's degrees in biology were earned by
women \cite{aps_bachelors_2022} while the
previously discussed Irish dataset found only 54\% of male students but
78\% of female students completed Biology at the end of secondary
school \cite{delaney_understanding_2019}. We found that in 2018, 12U Biology
had a median female proportion of \(66.7 \pm 0.4\%\), a gap between male
and female students three percentage points greater than what is seen in
12U Physics. And although physics is narrowing the gender gap, albeit
slowly, the gender gap in biology has been expanding; the average yearly
increase in the median female proportion was 3.8 times higher for 12U
Biology than for 12U Physics. Previous work looking at Ontario data
tracked the cohort of students entering Grade 9 in 2005 and calculated
the percentage of these students who took a Grade 12 science course
within the next 5 years. Their results found 25.7\% of female and 15.3\%
of male students took biology while 9.1\% of female and 16.6\% of male
students took physics \cite{card_high_2021}. These rates are all smaller
than our calculated average $SCR$s for these courses (these are not
equivalent but are analogous measures). This reinforces our overall
conclusions of growing interest in STEM while showing differing rates of
growth between male and female students.

Through a linear mixed effects model calculating the yearly average
change in Student Continuation Rates, we examined what has been driving
these changing gender imbalances. In biology, change is occurring in
both directions. Compared with a decade earlier, a significantly higher
proportion of women are now continuing from grade 10 science to 12U
Biology while significantly fewer men are (\(+ 0.75 \pm 0.08\) and
\(- 0.44 \pm 0.08\) percentage points per year, respectively). This, we
believe, is evidence of a positive feedback loop; when the percentage of
female students in biology classrooms increases, other female students
experience a greater sense of belonging in these spaces, leading to a
rise in enrolment. The corollary is that male students may then~perceive
these settings as less welcoming, resulting in decreasing participation.

If educators and researchers are not mindful of this new, widening
division, in a few decades we will face the exact same issue in
biology-related STEM fields as we currently do in physics, engineering,
and computer science. As a result, the same problems that stem from the
lack of gender diversity, e.g., decreased productivity, a dearth of new
ideas, and human rights issues from bias or a lack of equitable
opportunities may also begin to manifest themselves in the biological
sciences.

\section{Conclusions}\label{conclusions}

These results suggest there have been great strides made to increase
enrolment in STEM, particularly for women. However, we have also shown
many issues in equitable participation in STEM education still exist
including the large attrition of all students between grade 10 science
and grade 12 physics, the negligible progress made to close the existing
gender gap in physics, and the now widening gender gap in biology. As
researchers and educators have developed a growing understanding of how
to design successful initiatives to promote STEM education, particularly
for women, we believe a second generation of more targeted interventions
is now required. As such, we have four recommendations based on our
results.

\begin{enumerate}
\def\labelenumi{\arabic{enumi}.}
\item
  Our results add to a growing literature that has compared enrolment
  trends and gender gaps across different STEM disciplines \cite{cheryan_why_2017}. The results are clear: there are large discrepancies across different STEM fields and the tendency for researchers,
  educators, of governments to talk about STEM as one subject hides this
  fact. This hinders progress to address existing challenges such as the
  large persistent gender gap in physics and engineering; a problem
  which needs to be corrected to properly meet the rising demands for
  STEM-educated individuals in the 21\textsuperscript{st} century.
\item
  Similar to the flaw in conceptualizing STEM as a single entity, women
  are not one homogeneous group. As briefly summarized when explaining
  the intent of this work (sec. 1c), examining the intersectionality of
  gender with other demographic factors such as race or socioeconomic
  status will provide a clearer picture of who does or does not choose
  to participate in STEM education. A subsequent research project using
  these data examines the intersectionality of gender and other
  demographic factors for Ontario schools (Corrigan et al., in
  preparation) and we encourage other researchers to also consider these
  important distinctions in their future work.
\item
  Access to high-quality enrolment data related to gendered enrolment in
  secondary schools is currently not widely available. As we have
  outlined, this period is pivotal in shaping the future education and
  employment opportunities of students and a lack of transparency at
  this level acts as a hindrance to organizations seeking to improve
  equitable outcomes for STEM education. It would also help flag
  enrolment trends such as the growing disparity between male and female
  students in biology we have discovered through this analysis.
\item
  Future initiatives or interventions to improve or promote STEM
  education need to be designed with a nuanced consideration of both
  STEM and gender. Without clearly targeted efforts, e.g., an
  intervention to promote women in high school physics instead of a more
  general intervention to promote women in STEM, we will not ameliorate
  the persistent lack of diversity that hinders progress in these
  fields.
\end{enumerate}

\appendix

\section*{Appendix A -- Regression Results }\label{apdx:reg_results}

\begin{table} [h!]
\caption{\label{tab:delta_SCR_reg} 
Mixed effects linear regression estimates predicting yearly change in SCR for senior STEM courses.}
\begin{ruledtabular}
\begin{tabular}{lddd}
 & \multicolumn{3}{c}{Female Students}\\
 \cline{2-4}\\
 \vspace{-2em} \\
Course & \multicolumn{1}{r}{Estimate} & \multicolumn{1}{l}{Standard Error} & \multicolumn{1}{r}{$p$-value} \\
\hline
11U Biology & -0.340 & 0.061 & <.001 \\
11U Chemistry & +0.640 & 0.062 & <.001 \\
11U Physics & +0.505 & 0.053 & <.001 \\
12U Biology & -0.440 & 0.084 & <.001 \\
12U Chemistry & +0.417 & 0.079 & <.001 \\
12U Physics & +0.555 & 0.060 & <.001 \\
12U Functions & -0.192 & 0.079 & .015 \\
12U Calculus & +0.304 & 0.065 & <.001 \\
\hline
\vspace{-1em}\\
& \multicolumn{3}{c}{Male Students}\\
 \cline{2-4}\\
 \vspace{-2em} \\
Course & \multicolumn{1}{r}{Estimate} & \multicolumn{1}{l}{Standard Error} & \multicolumn{1}{r}{$p$-value} \\
\hline
11U Biology & +0.583 & 0.059 & <.001 \\
11U Chemistry & +1.114 & 0.061 & <.001 \\
11U Physics & +0.720 & 0.055 & <.001 \\
12U Biology & +0.749 & 0.078 & <.001 \\
12U Chemistry & +1.037 & 0.077 & <.001 \\
12U Physics & +0.490 & 0.069 & <.001 \\
12U Functions & +0.144 & 0.079 & .066 \\
12U Calculus & +0.383 & 0.067 & <.001 \\
\hline
\vspace{-1em}\\
& \multicolumn{3}{c}{Interaction Terms (Female as Reference)}\\
 \cline{2-4}\\
 \vspace{-2em} \\
Course & \multicolumn{1}{r}{Estimate} & \multicolumn{1}{l}{Standard Error} & \multicolumn{1}{r}{$p$-value} \\
\hline
11U Biology & -0.922 & 0.072 & <.001 \\
11U Chemistry & -0.473 & 0.071 & <.001 \\
11U Physics & -0.215 & 0.062 & <.001 \\
12U Biology & -1.189 & 0.087 & <.001 \\
12U Chemistry & -0.620 & 0.081 & <.001 \\
12U Physics & +0.065 & 0.075 & .387 \\
12U Functions & -0.336 & 0.100 & <.001 \\
12U Calculus & -0.079 & 0.082 & .333 \\
\end{tabular}
\end{ruledtabular}
\end{table}

\bibliography{references}
\end{document}